\documentclass[superscriptaddress,showpacs,aps,pra,twocolumn,amsmath,amssymb]{revtex4-1}
\pdfoutput=1
\usepackage{times}
\usepackage{hyperref}
\usepackage{xspace}
\usepackage{graphicx}
\usepackage{hypernat,xspace}
\usepackage[figure,table]{hypcap}

\hypersetup{colorlinks=true,
linkcolor=blue,
filecolor=blue,
citecolor=blue,
urlcolor=blue,
pdfauthor={Yongyong Cai, Matthias Rosenkranz, Zhen Lei, Weizhu Bao},
pdftitle={Mean-field regime of trapped dipolar Bose-Einstein condensates in one
  and two dimensions},
pdfkeywords={dipolar BEC, groundstate}}

\mathchardef\ordinarycolon\mathcode`\:
\mathcode`\:=\string"8000
\begingroup \catcode`\:=\active
  \gdef:{\mathrel{\mathop\ordinarycolon}}
\endgroup

\newcommand{\BEC}{BEC\xspace}
\newcommand{\GPE}{GPE\xspace}
\newcommand{\eg}{\textit{e.g.}\xspace}
\newcommand{\ie}{\textit{i.e.}\xspace}
\newcommand{\di}{\ensuremath{d}}
\newcommand{\im}{\ensuremath{i}}
\newcommand{\eu}{\ensuremath{e}}
\newcommand{\bvec}[1]{\ensuremath{\mathbf{#1}}}
\DeclareMathOperator{\erfc}{erfc}

\begin{document}
\title{Mean-field regime of trapped dipolar Bose-Einstein
  condensates in one and two dimensions}
\author{Yongyong Cai}
\author{Matthias Rosenkranz}
\email{matrm@nus.edu.sg}
\affiliation{Department of Mathematics, National University of
  Singapore, 119076, Singapore}
\author{Zhen Lei}
\affiliation{School of Mathematical Sciences, Fudan University,
  Shanghai 200433, China}
\affiliation{Key Laboratory of Nonlinear Mathematical Models and
Methods of Ministry of Education, Fudan University, Shanghai 200433,
China}
\affiliation{Shanghai Key Laboratory for Contemporary Applied
Mathematics, Fudan University, Shanghai 200433, China}
\author{Weizhu Bao}
\email{bao@math.nus.edu.sg}
\homepage{http://www.math.nus.edu.sg/~bao/}
\affiliation{Department of Mathematics, National University of
  Singapore, 119076, Singapore}
\affiliation{Center for Computational Science and Engineering,
  National University of Singapore, 117543, Singapore}


\date{\today}

\begin{abstract}
  We derive rigorous one- and two-dimensional mean-field equations for
  cigar- and pancake-shaped dipolar Bose-Einstein condensates with
  arbitrary polarization angle.  We show how the dipolar interaction
  modifies the contact interaction of the strongly confined atoms.  In
  addition, our equations introduce a nonlocal potential, which is
  anisotropic for pancake-shaped condensates.  We propose to observe
  this anisotropy via measurement of the condensate aspect ratio.  We
  also derive analytically approximate density profiles from our
  equations.  Both the numerical solutions of our reduced mean-field
  equations and the analytical density profiles agree well with
  numerical solutions of the full Gross-Pitaevskii equation while
  being more efficient to compute.
\end{abstract}
\maketitle

\section{Introduction}

Quantum-degenerate gases with long-range interactions have received
much attention recently both from experimental and theoretical
studies.  In conventional experiments with bosonic quantum gases,
short-range interactions have played a leading role and are well
described by the s-wave scattering length~\cite{PitStr03}.  With the
realization of a dipolar chromium Bose-Einstein condensate (\BEC) it
is now possible to go beyond such isotropic interactions in degenerate
gases~\cite{GriWerHen05,StuGriKoc05}.  Dipolar interactions have a
long-range and anisotropic component.  These features crucially affect
the ground state properties~\cite{GorRzaPfa00,YiYou00},
stability~\cite{SanShlZol00,SanShlLew03}, and dynamics of the
gas~\cite{YiYou01}.  Furthermore, they offer a route for studying
exciting many-body quantum effects such as a superfluid-crystal
quantum phase transition~\cite{BueDemLuk07}, supersolids
~\cite{GorSanLew02} or even topological order~\cite{MicBreZol06} (for
a review of the experimental and theoretical progress in dipolar gases
see Refs.~\cite{LahMenSan09,Bar08}).

\textsuperscript{52}Cr possess a comparatively large magnetic dipole
moment of $6$ Bohr magnetons.  A large magnetic dipole moment makes
atomic \BEC{}s an ideal candidate for studying the interplay between
contact and dipole-dipole interactions.  By reducing the s-wave
scattering length via a Feshbach resonance it is even possible that
dipole-dipole interactions dominate the properties of the
\BEC~\cite{KocLahMet08,FatRoaDei08}.  Dipolar effects have also been
observed in a spinor alkali condensate~\cite{VenLesGuz08}.
Furthermore, dc electric fields can induce large electric dipole
moments in alkali atoms~\cite{MarYou98}.  Systems with a large
permanent electric dipole moment include heteronuclear
molecules~\cite{HaiKleBha04,SagSaiBer05}, which are harder to cool to
quantum degeneracy~\cite{NiOspMir08,OspPeeNi08}, and Rydberg
atoms~\cite{VogVitZha06}.

\begin{figure}[bB]
  \centering
  \includegraphics[width=.8\linewidth]{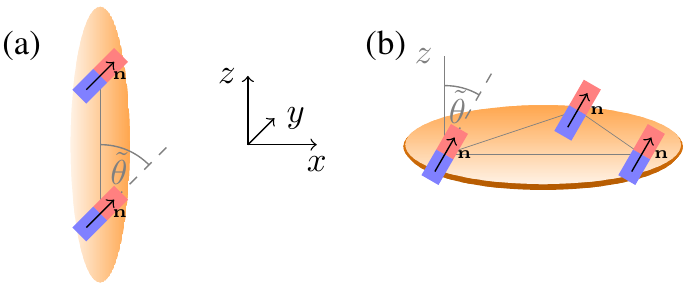}
  \caption{(Color online) In the quasi-1D setup in (a) the dipolar
    \BEC is confined to the $z$ direction.  In the quasi-2D setup in
    (b) the atoms are confined to the $x$-$y$ plane.  The dipoles are
    polarized along the axis $\bvec n = (n_x,n_y,n_z)$ with polar
    angle $\tilde\theta$ (\ie, $n_z =
    \cos\tilde\theta$).}\label{fig:setup}
\end{figure}

Complementing the tremendous experimental progress, many properties of
trapped dipolar \BEC{}s have been investigated theoretically.  The
Gross-Pitaevskii equation (\GPE) with a nonlocal potential determines
its ground state density profile in three dimensions and at zero
temperature ~\cite{YiYou00}.  Neglecting the kinetic energy term,
\citet{ODeGioEbe04,EbeGioODe05} have shown that, remarkably, the
ground state density profile of a harmonically trapped
three-dimensional (3D) dipolar \BEC is an inverted parabola just as in
the nondipolar \BEC.  On the other hand, in most experiments to date a
strong harmonic trap (or optical lattice) along one or two axes
confines the dipolar \BEC to a cigar or pancake shape,
respectively~\cite{StuGriKoc05,FatRoaDei08}.  For these cases,
\citet{ParODe08} have derived one- (1D) and two-dimensional (2D)
density profiles of a dipolar \BEC both in the Thomas-Fermi (TF) limit
and the 1D and 2D mean-field limit.  However, their results are only
valid for polarization along the symmetry axis.  Several authors have
derived effective dipolar potentials in lower dimensions for either
axial or transverse
polarizations~\cite{GioODe04,SinSan07,PedSan05,NatPedSan09,YiPu06,KomCoo07}.
To our knowledge, an effective dipolar potential valid for arbitrary
polarization direction of quasi-1D and quasi-2D dipolar \BEC{}s has
not been proposed yet.

Apart from a conceptual clarity, effective equations for lower
dimensional dipolar \BEC{}s also offer a clear advantage for numerical
computations.  For strong trap anisotropies the time scales along the
compressed and elongated axes are very different, which makes an
accurate numerical treatment hard.  Instead of solving the full 3D
problem, it is hence desirable to find governing equations for lower
dimensional dipolar \BEC{}s which are suitable for efficient numerical
methods.  This is particularly important for very strong confinement
or low densities, where the usual TF approximation for the full 3D
\GPE becomes invalid~\cite{PitStr03}.

In this paper, we present mean-field equations for trapped dipolar
\BEC{}s in one and two dimensions polarized along an arbitrary axis.
Our equations are based on a mathematically rigorous dimension
reduction of the 3D Gross-Pitaevskii equation (\GPE) to lower
dimensions.  For \BEC{}s without dipolar interactions the formal
analysis~\cite{BaoTan03,BaoJakMar03,Luca09} and rigorous
analysis~\cite{NaoMeScWe05,NaoCasMe08,LieSeiYng03, YngLieSei04} of
such a dimension reduction has been discussed extensively in the
physical and mathematical literature.  On the other hand, only few
mathematically rigorous results are available for the dimension
reduction of dipolar \BEC{}s, such as Ref.~\cite{CarMarSpa08}.  The
main advantages of our effective equations over previously derived
results are: (i) valid for arbitrary dipole alignment, (ii) well
amenable for efficient numerical computations typically based on the
Fourier transformation and related methods.

For the derivation of the 1D (2D) case we assume that the \BEC is in
the ground state of the radial (axial) trap (see
Fig.~\ref{fig:setup}).  We find that the ground state of the
lower-dimensional dipolar \BEC is determined by a modified contact
interaction term and a nonlocal potential.  Compared to a confined
\BEC with only s-wave interactions here the contact interaction also
depends on the strength of the dipolar interaction and the
$z$-component of the polarization axis.  Crucially, it is independent
of the transverse components of the polarization.  In 2D the nonlocal
term introduces anisotropy in the \BEC ground state density, which can
be measured, \eg, in time-of-flight experiments as a modified aspect
ratio of the \BEC~\cite{StuGriKoc05,GioGoePfa03}.  We discuss the
aspect ratio of the \BEC for the intermediate regime between the
special cases of parallel and orthogonal polarization, which have been
studied before~\cite{ParODe08,ODeGioEbe04}.  Furthermore, we present
simple analytical density profiles in 1D and 2D derived from our
mean-field equations.  We compare ground states of the quasi-1D and
quasi-2D \BEC at different polarizations with the ground states of the
full 3D \BEC and find good agreement.  In particular, our ground
states are a good approximation to the ground states of the full \GPE
in regimes where the TF approximation fails.  In the limit of strong
axial confinement we can cast our nonlocal potential into a form
similar to the Poisson equation found for 3D dipolar
\BEC{}s~\cite{ODeGioEbe04}.

This paper is organized as follows. In Sec.~\ref{sec:3d}, we introduce
the model of a 3D dipolar \BEC at zero temperature.  As our first main
result, in Sec.~\ref{sec:1d} we present a mean-field equation for a
quasi-1D dipolar \BEC.  We compare the ground state solutions of this
1D equation with the full 3D computation and an approximate analytical
solution.  In Sec.~\ref{sec:2d}, we present our second main result,
namely, a mean-field equation for a quasi-2D dipolar \BEC.  Again we
compare its solutions to the 3D \GPE solution and our analytical
approximation.  Moreover, we calculate the aspect ratio of the \BEC if
the polarization of the dipoles is changed continuously from the
longitudinal to a radial axis.  We conclude in
Sec.~\ref{sec:conclusion}.  In Appendix~\ref{app:1d} and \ref{app:2d}
we present details of the dimension reduction from the 3D \GPE to the
1D and 2D mean-field equations, respectively.  In
Appendix~\ref{app:potentials} we derive closed forms for the nonlocal
potentials in 1D and 2D for arbitrary dipole alignment.

\section{3D model}\label{sec:3d}
We consider a dilute dipolar \BEC at zero temperature trapped in a
harmonic potential $V(\bvec r) = \tfrac{m}{2} (\omega_x^2 x^2 +
\omega_y^2 y^2 + \omega_z^2 z^2)$, where $m$ is the particle mass and
$\omega_{x,y,z}$ are the trap frequencies.  We focus on atomic \BEC{}s
with a magnetic dipole moment but it is straightforward to extend the
analysis to degenerate bosonic gases with electric dipole moments.  We
assume that the atoms are polarized along a dipolar axis $\bvec n =
(n_x, n_y, n_z)$ with $\sum_i n_i^2 = 1$.  Away from shape resonances,
the wave function $\psi(\bvec r, t)$ of the gas is governed by the
GPE~\cite{MarYou98,YiYou00,DebYou01}
\begin{equation}\label{eq:GPE}
  \im\hbar\partial_t \psi(\bvec r, t) = \left[ -\frac{\hbar^2}{2m}
    \nabla^2 + V(\bvec r) + g|\psi|^2 + \Phi_{dd} \right] \psi(\bvec r,
  t),
\end{equation}
where $g = 4\pi\hbar^2 a_s/m$ is the contact interaction strength with
s-wave scattering length $a_s$.  The dipolar potential $\Phi_{dd}$ is
given by the convolution
\begin{equation}\label{eq:Phi_dd}
  \Phi_{dd} = \int \di^3\bvec r' U_{dd}(\bvec r - \bvec r') |\psi(\bvec r', t)|^2
\end{equation}
with the dipole interaction
\begin{equation}\label{eq:U_dd}
  U_{dd}(\bvec r) = \frac{C_{dd}}{4\pi} \frac{1 -
    3\cos^2\theta}{|\bvec r|^3}.
\end{equation}
Here, $\theta$ is the angle between the polarization axis $\bvec n$
and the relative position of two atoms (\ie, $\cos\theta = \bvec n
\cdot \bvec r/|\bvec r|$).  For magnetic dipoles we have $C_{dd} =
\mu_0 \mu_d^2$, where $\mu_0$ is the magnetic vacuum permeability and
$\mu_d$ the dipole moment, and for electric dipoles we have $C_{dd} =
d^2/\epsilon_0$, where $\epsilon_0$ is the vacuum permittivity and $d$
the electric dipole moment.  We note that it is possible to modify the
dipolar interaction $C_{dd}$ by means of a rotating magnetic
field~\cite{GioGoePfa02}.

We use a mathematical identity to write the dipole interaction
Eq.~\eqref{eq:U_dd} as~\cite{ODeGioEbe04,BaoCaiWan10}
\begin{equation}\label{eq:U_dd2}
  U_{dd}(\bvec r) = -C_{dd} \left( \frac{1}{3} \delta(\bvec r)
    + \partial_{\bvec n\bvec n} \frac{1}{4\pi |\bvec r|} \right).
\end{equation}
Here we denote with $\partial_{\bvec n} = n_x\partial_x +
n_y\partial_y + n_z\partial_z$ the derivative along the dipole axis
and $\partial_{\bvec n\bvec n} = \partial_{\bvec n} (\partial_{\bvec
  n})$.  Inserting Eq.~\eqref{eq:U_dd2} and Eq.~\eqref{eq:Phi_dd} into
the \GPE, Eq.~\eqref{eq:GPE}, results in
\begin{equation}\label{eq:psi_3d_dim}
  i\hbar \partial_t \psi = \left[ -\frac{\hbar^2}{2m}
    \nabla^2 + V(\bvec r) + \left( g - \frac{C_{dd}}{3}
    \right)|\psi|^2 + \tilde\Phi_\text{3D} \right] \psi.
\end{equation}
We note that two terms contribute to the dipolar interaction.  The
first term in Eq.~\eqref{eq:U_dd2} reduces the contact interaction
strength [third term in Eq.~\eqref{eq:psi_3d_dim}], while the second
term in Eq.~\eqref{eq:U_dd2} contributes the potential
$\tilde\Phi_\text{3D} = -C_{dd} \partial_{\bvec n\bvec n} \int d^3
\bvec r' U_\text{3D}(\bvec r - \bvec r') |\psi(\bvec r', t)|^2$ with
kernel $U_\text{3D}(\bvec r) = 1/4\pi |\bvec r|$.

We introduce dimensionless quantities by rescaling lengths with $\bvec
r \rightarrow \bvec r a_0$, times with $t \rightarrow t/\omega_0$,
energies with $\hbar\omega_0$, and the wave function with $\psi
\rightarrow \psi \sqrt{N/a_0^3}$, where $\omega_0$ is the smallest
trap frequency in the system [$\omega_0 = \min(\omega_x, \omega_y,
\omega_z)$], $a_0 = \sqrt{\hbar/m\omega_0}$ the corresponding magnetic
length, and $N$ the total number of atoms in the \BEC.  After
rescaling, in dimensionless form Eq.~\eqref{eq:psi_3d_dim} is given by
\begin{subequations}
  \begin{align}
    i\partial_t \psi &= \left[-\frac{1}{2} \nabla^2 + V(\bvec r) +
      \beta(1 - \epsilon_{dd}) |\psi|^2 + \Phi_\text{3D} \right] \psi,\label{eq:psi_3d_a}\\
    \Phi_\text{3D} &= -3\beta\epsilon_{dd} \partial_{\bvec n\bvec n}
    \int d^3 \bvec r' U_\text{3D}(\bvec r - \bvec r') |\psi(\bvec r',
    t)|^2.\label{eq:phi_3d}
  \end{align}
\end{subequations}
where $\beta = 4\pi N a_s/a_0$ and $\epsilon_{dd} = C_{dd}/3g$ defines
a natural dimensionless parameter for the relative strength of dipolar
and s-wave interactions.  The dimensionless trapping potential is
$V(\bvec r) = \frac{1}{2} (\gamma_x^2 x^2 + \gamma_y^2 y^2 +
\gamma_z^2 z^2)$ with $\gamma_x = \omega_x/\omega_0$, $\gamma_y =
\omega_y/\omega_0$, $\gamma_z = \omega_z/\omega_0$.

\section{Quasi-1D dipolar BEC}\label{sec:1d}
By choosing a sufficiently large radial trap frequency it is possible
to freeze the radial motion of the \BEC~\cite{GoeVogLea01}.  If the
extent of the radial cloud is much larger than the s-wave scattering
length, this is the limit of a quasi-1D \BEC~\cite{PetShlWal00}.  In
Fig.~\ref{fig:setup}(a) we illustrate the geometry of this setup.  In
this section, we present an intuitive mean-field equation for the
axial wave function of such a strongly confined dipolar \BEC{}.
Our equation is based on a reduction of the 3D \GPE to 1D assuming a
strong radial confinement.

\subsection{1D mean-field equation}
In order to derive a mean-field equation for the axial wave function
of the condensate we assume that $\omega_z \ll \omega_x = \omega_y =:
\omega_\perp$ and $gn_0 \ll \hbar\omega_\perp$, where $n_0$ is the
peak density of the BEC.  Moreover, we require that dipole
interactions do not excite radial modes, \ie, $\Phi_{dd} \ll
\hbar\omega_\perp$.  Then the radial modes of the \BEC are in the
ground state of the transversal harmonic trap and the order parameter
$\psi$ of the \BEC factorizes.  We write the factorized wave function
as
\begin{align}\label{eq:factorization_1d}
  \psi(\bvec r, t) &= \eu^{-\im\omega_\perp t} w_\text{2D}(x, y)
  \psi_\text{1D}(z, t),\\
  w_\text{2D}(x, y) &= \sqrt{\frac{m\omega_\perp}{\pi\hbar}}
  \eu^{-m\omega_\perp(x^2+y^2)/2\hbar}.
\end{align}
In this section we rescale equations in terms of the dimensionless
lengths $\bvec r \rightarrow \bvec r a_z$, times $t \rightarrow
t/\omega_z$, and the axial wave function $\psi_\text{1D}
\rightarrow \psi_\text{1D} \sqrt{N/a_z}$ with $a_z =
\sqrt{\hbar/m\omega_z}$ the magnetic length in $z$ direction.
Energies are expressed in units of $\hbar\omega_z$.  Given these
assumptions, in Appendix~\ref{app:1d} we show that the 3D \GPE,
Eq.~\eqref{eq:GPE}, reduces to an equation for the axial wave
function $\psi_\text{1D}$
\begin{widetext}
\begin{subequations}\label{eq:psi_1d}
\begin{align}
  \im\partial_t \psi_\text{1D}(z, t) &= \left\{
    -\frac{1}{2} \partial_{zz} + V_\text{1D}(z) +
    \frac{\beta_\text{1D}\gamma}{2\pi}\left[ 1 +
      \frac{\epsilon_{dd}}{2} \left(1 - 3n_z^2 \right) \right]
    |\psi_\text{1D}(z, t)|^2 + \Phi_\text{1D}
  \right\} \psi_\text{1D}(z, t),\label{eq:psi_1d_a}\\
  \Phi_\text{1D} &= \frac{3\beta_\text{1D}\epsilon_{dd}
    \sqrt\gamma}{8\sqrt{2\pi}} \left(1 - 3n_z^2 \right) \partial_{zz}
  \int_{-\infty}^\infty \di z' U_\text{1D} (z-z') |\psi_\text{1D}(z',
  t)|^2,\label{eq:phi_1d}
\end{align}
\end{subequations}
\end{widetext}
where $V_\text{1D}(z) = z^2/2$, $\beta_\text{1D} = gN/\hbar\omega_z
a_z^3 = 4\pi N a_s/a_z$ and $n_z$ is the $z$ component of the dipole
axis $\bvec n$.  The trap aspect ratio is given by $\gamma =
\omega_\perp/\omega_z$.  We find for the kernel $U_\text{1D}$
\begin{equation}\label{eq:U_1d}
  U_\text{1D}(z) = \eu^{\gamma z^2/2} \erfc(|z|\sqrt{\gamma/2}),
\end{equation}
where $\erfc$ is the complementary error function.  In
Appendix~\ref{app:potentials} we calculate the derivative in
Eq.~\eqref{eq:phi_1d} and give a closed form of the resulting
convolution integral.  However for our focus on numerical computation
of the ground state, expression~\eqref{eq:phi_1d} is better suited.
It is worth pointing out that the only approximation in the derivation
of Eq.~\eqref{eq:psi_1d} from the \GPE is the factorization in
Eq.~\eqref{eq:factorization_1d} with the choice of a Gaussian as the
radial wave function $w_\text{2D}(x,y)$.

\begin{figure}
  \centering%
  \includegraphics[width=.8\linewidth]{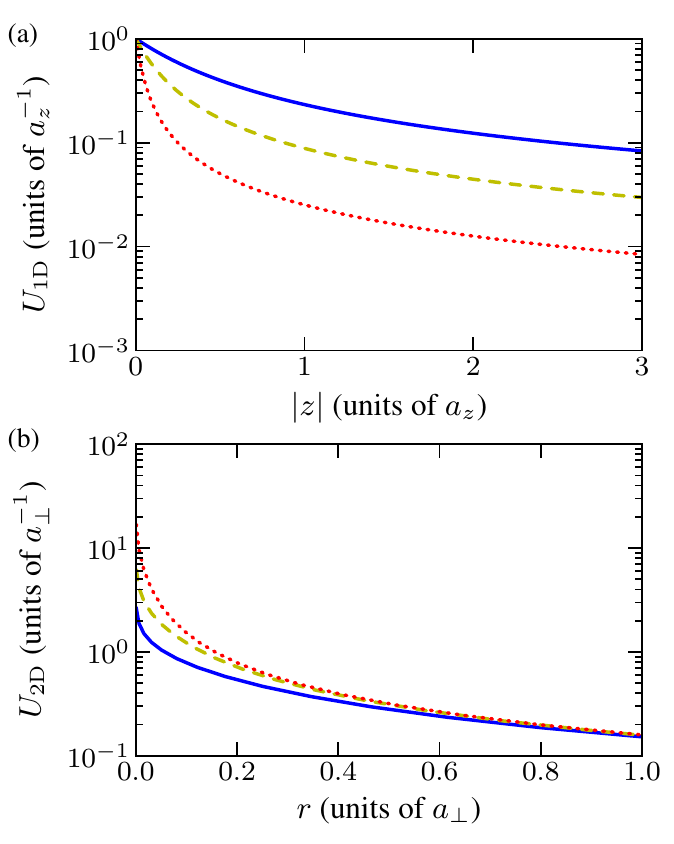}
  \caption{(a) The kernel $U_\text{1D}$, Eq.~\eqref{eq:U_1d}, for
    $\gamma = 10$ (solid line), $80$ (dashed), $1000$ (dotted).  (b)
    $U_\text{2D}$, Eq.~\eqref{eq:U_2D}, for $\gamma = 1/10$ (solid),
    $1/80$ (dashed), $1/1000$ (dotted).  In both cases increasing the
    confinement of the \BEC leads to increasingly local
    behavior.}\label{fig:U}
\end{figure}

The formula for the wave function of a quasi-1D dipolar \BEC,
Eq.~\eqref{eq:psi_1d_a}, is very intuitive.  It has the same structure
as the corresponding 3D expression Eq.~\eqref{eq:psi_3d_a}.  Notably,
the effect of the dipolar interaction in 1D is an altered contact
interaction strength and the introduction of a nonlocal potential
$\Phi_\text{1D}$.  In Fig.~\ref{fig:U}(a) we plot the kernel
$U_\text{1D}$ of the potential, Eq.~\eqref{eq:U_1d}, for different
trap aspect ratios.  In contrast to the kernel $U_\text{3D}$ in 3D,
the kernel of the 1D potential does not diverge at the origin.
Instead we find $U_\text{1D}(z) = 1 - \sqrt{2\gamma} |z|/\sqrt\pi +
\mathcal{O}(|z|^2)$ for $|z| \rightarrow 0$.  In the opposite limit,
$|z| \rightarrow \infty$, we find $U_\text{1D}(z) \sim
\sqrt{2}/\sqrt{\pi\gamma} |z|$, \ie, it scales with $1/|z|$ as the 3D
kernel.  These properties allow for efficient numerical methods based
on Eq.~\eqref{eq:psi_1d}~\cite{BaoCaiWan10}.

\subsection{Ground state}
The third term in Eq.~\eqref{eq:psi_1d_a} describes the altered
contact interaction, which now depends on the dipolar interaction
strength $\beta_\text{1D}\epsilon_{dd}$.  The anisotropy of the
dipolar interaction is manifest in this term in the dependence on the
$z$ component $n_z = \cos\tilde\theta$ of the dipole axis, where
$\tilde\theta$ is the angle between dipole and $z$ axis
[cf. Fig.~\ref{fig:setup}(a)].  If the dipoles are aligned along the
longitudinal \BEC axis ($n_z = 1$), the effective contact interaction
reduces by a factor of $(1 - \epsilon_{dd})$.  Neglecting the nonlocal
part of Eq.~\eqref{eq:psi_1d_a} this results in a reduced \BEC length.
If the dipole axis is perpendicular to the \BEC axis ($n_z = 0$), the
\BEC length increases since the contact interaction is larger by a
factor $(1 + \epsilon_{dd}/2)$.  Intuitively, these two cases can be
understood in terms of a string of magnets: magnets with poles aligned
head to tail attract each other, while magnets in a head to head or
tail to tail configuration repel each other.  However, our equation
also maps out all intermediate configurations between these two
special cases.  Finally, we notice the familiar increase of the
effective s-wave scattering strength in 1D by a factor $\gamma/2\pi$,
which is due to the choice of a Gaussian ground state in the
factorization
Eq.~\eqref{eq:factorization_1d}~\cite{Ols98,BaoJakMar03}.

The potential $\Phi_\text{1D}$ in Eq.~\eqref{eq:psi_1d_a} describes
the nonlocal effect of the dipolar interaction on the \BEC.  The shape
of the kernel $U_\text{1D}$ [see Fig.~\ref{fig:U}(a)] reveals that
this potential becomes more local with increasing trap aspect ratio
$\gamma$.  Moreover, for large $\gamma$ we expect that
$\Phi_\text{1D}$ does not affect the shape of the \BEC significantly.
Owing to the properties of convolutions we may apply the second
derivative in $\Phi_\text{1D}$ only to the density
$|\psi_\text{1D}|^2$ in the integral.  However, for the ground state
we expect this density to become flatter for large $\gamma$ so that
the derivative becomes smaller.  Since the contact interaction scales
linearly with $\gamma$, the contact term dominates over the nonlocal
potential.  Similar to the modified contact interaction our
Eq.~\eqref{eq:phi_1d} explicitly states the dependence of
$\Phi_\text{1D}$ on the $z$-component of the dipole axis (and predicts
no dependence on other components).

We have seen that the dipolar interaction in 1D is composed of a local
(or contact) interaction and a nonlocal interaction.  In order to
determine the sign of the nonlocal interaction, we evaluate the
corresponding energy via Fourier transformation.  The energy is given
by $\tfrac{1}{2} \int dz \Phi_\text{1D} |\psi_\text{1D}(z, t)|^2 =
\tfrac{1}{2} \int dk_z \widehat{\Phi_\text{1D}}
\widehat{|\psi_\text{1D}|^2}^*(k_z, t) = -\tfrac{3\beta_\text{1D}
  \epsilon_{dd} \sqrt\gamma}{16} (1-3n_z^2) \int dk_z k_z^2
\widehat{U_\text{1D}}(k_z) \left|\widehat{|\psi_\text{1D}|^2}(k_z,
  t)\right|^2$, where $\widehat{f}(k_z) = (1/\sqrt{2\pi}) \int dz f(z)
\eu^{-\im k_z z}$ denotes the Fourier transform of a function $f$.
Since the Fourier transforms $\widehat{U_\text{1D}}$ and
$\bigl|\widehat{|\psi_\text{1D}|^2}\bigr|^2$ are positive (see
Appendix~\ref{app:1d}), the sign of the nonlocal potential is opposite
the sign of the modification in the contact term
$\tfrac{\beta_\text{1D}\epsilon_{dd}\gamma}{4\pi}(1-3n_z^2)$.  This
means that the nonlocal potential counteracts the action of the
contact term: if the contact term is repulsive, the nonlocal potential
becomes attractive and vice versa.

Another useful observation is the following.  The \BEC ground state is
not affected by the presence of dipolar interactions for the ``magic
angle'' $\cos\tilde\theta_m = 1/\sqrt{3}$~\cite{Meh83}.  In this case,
the 3D dipolar potential $U_{dd}$ vanishes and accordingly
Eq.~\eqref{eq:psi_1d} reduces to the \GPE for a quasi-1D \BEC without
dipolar interaction.  This observation could be useful for very
sensitive matter wave interferometers, where the dipole interaction
dominates the decoherence when the s-wave scattering length has been
reduced via a Feshbach resonance~\cite{FatRoaDei08}.

We can derive an analytical solution for the density of the quasi-1D
dipolar \BEC if the contact interaction term in
Eq.~\eqref{eq:psi_1d_a} is repulsive and dominates the
dynamics~\cite{PitStr03}.  In this case, we neglect the kinetic and
nonlocal parts in Eq.~\eqref{eq:psi_1d_a}.  Assuming a stationary
solution, $\psi_\text{1D}(z, t) = \eu^{-\im\mu_z t}
\sqrt{n_\text{1D}(z)}$, we find the density profile $n_\text{1D}(z) =
[\mu_z - (z^2/2)] \{\beta_\text{1D}\tfrac{\gamma}{2\pi} [1 +
\tfrac{\epsilon_{dd}}{2} (1 - 3n_z^2)]\}^{-1}$, where $\mu_z$ is the
chemical potential along $z$.  The density vanishes for $z \geq Z =
\sqrt{2\mu_z}$.  We obtain the halfwidth of the condensate $Z$ by
evaluating the normalization condition $\int \di z n(z) = 1$, which
yields
\begin{equation}
  Z = \left\{ \frac{3}{2} \frac{\beta_\text{1D}\gamma}{2\pi} \left[ 1 +
      \frac{\epsilon_{dd}}{2} (1 - 3n_z^2) \right] \right\}^{1/3}.
\end{equation}
Inserting this expression into the density profile gives
\begin{equation}\label{eq:n_TF}
  n_\text{1D}(z) = \frac{3}{4Z} \left(1 - \frac{z^2}{Z^2} \right).
\end{equation}
\citet{ParODe08} have derived analytical 1D densities for the special
case of dipoles aligned along the $z$ axis.  In this case, we note
that our generalized expression Eq.~\eqref{eq:n_TF} coincides with
their density in the limit termed ``1D mean-field regime''.
Generalizing the criterion for the validity of the mean-field regime
in Refs.~\cite{ParODe08,MunDel08}, we see that Eq.~\eqref{eq:n_TF} is
a good approximation for the axial density profile of an elongated
dipolar \BEC if
\begin{equation}
  \frac{\beta_\text{1D}}{4\pi\sqrt\gamma} \left[1 +
    \frac{\epsilon_{dd}}{2} (1 - 3n_z^2) \right] \ll 1.
\end{equation}

\begin{figure}
  \centering
  \includegraphics[width=\linewidth]{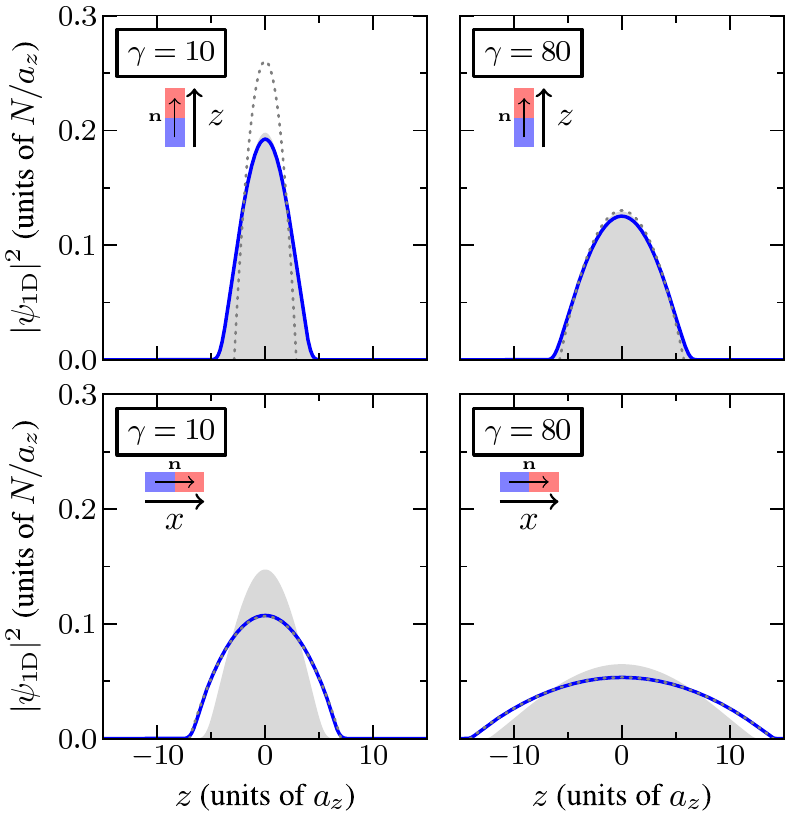}
  \caption{(Color online) Linear density of the quasi-1D \BEC
    according to the solution of our 2D equation,
    Eq.~\eqref{eq:psi_1d} (blue solid lines), the corresponding
    analytical prediction of Eq.~\eqref{eq:n_TF} (gray dotted), and
    the full 3D GPE of Eq.~\eqref{eq:GPE} (shaded area).  In
    the upper panel dipoles are aligned with the \BEC axis, while in
    the lower panel they are aligned perpendicular to the \BEC axis.
    We choose $\beta_\text{1D} = 100$, $\epsilon_{dd} = 0.9$ and the
    $\gamma$ given in the plots.}\label{fig:density_1d}
\end{figure}

In Fig.~\ref{fig:density_1d} we compare the density of the quasi-1D
\BEC determined via Eq.~\eqref{eq:psi_1d} with the analytical
prediction Eq.~\eqref{eq:n_TF} and the numerical solution for the full
3D \GPE in Eq.~\eqref{eq:GPE} after integrating over the transversal
directions.  If the dipole axis points perpendicular to the elongated
\BEC axis, we cannot distinguish the analytical density from the the
quasi-1D GPE result (see lower panel of Fig.~\ref{fig:density_1d}).
We have checked that for larger values of the trap aspect ratio
$\gamma$ the discrepancy between the full \GPE result and our
approximate 1D solution with perpendicular dipoles diminishes.  On the
other hand, if the dipoles are aligned with the \BEC axis our 1D
mean-field solution agrees very well with the solution of the full
\GPE (see upper panel of Fig.~\ref{fig:density_1d}).  Furthermore, we
note that the \BEC is compressed compared to the case with
perpendicular dipole axis.  This compression is caused by the
attraction of aligned dipoles in a 1D setup, and it is manifest in the
reduced contact interaction in Eq.~\eqref{eq:psi_1d_a} for $n_z=1$.
However, for axial dipoles the nonlocal term in
Eq.~\eqref{eq:psi_1d_a} produces an appreciable repulsive potential.
As a consequence, the \BEC is broadened as compared to the analytical
profile.  Therefore, in the regime of small or moderate interaction
energy $\beta_\text{1D}$, we see that the usual approach of neglecting
the kinetic and nonlocal terms is not sufficient to describe the
density profile.  On the other hand, our proposed 1D equation,
Eq.~\eqref{eq:psi_1d}, describes the \BEC accurately in the mean-field
regime at experimentally relevant trap aspect ratios $\gamma$.

\section{Quasi-2D dipolar BEC}\label{sec:2d}
In this section we consider a dipolar \BEC which is strongly confined
along the $z$ axis, \ie, $\gamma \ll 1$ [cf. Fig.~\ref{fig:setup}(b)].
Analogous to the preceding section, we assume that $gn_0 \ll
\hbar\omega_z$ and that the axial extend of the cloud is much
larger than the s-wave scattering length.  If the dipolar interactions
are also small compared to the axial trap energy
$\hbar\omega_z$, then the \BEC is in the ground state of the
axial harmonic trap.  This is the case of a quasi-2D \BEC,
where the \BEC wave function separates into
\begin{align}
  \psi(\bvec r, t) &= \eu^{-\im\omega_z t/2} \psi_\text{2D}(x, y, t)
  w_\text{1D}(z),\label{eq:factorization_2d}\\
  w_\text{1D}(z) &= \left( \frac{m\omega_z}{\pi \hbar} \right)^{1/4}
  \eu^{-m\omega_z z^2/2\hbar}.
\end{align}
In this section we use the dimensionless rescaling $\bvec r
\rightarrow \bvec r a_\perp$, $t \rightarrow t/\omega_\perp$, $\psi_\text{2D}
\rightarrow \psi_\text{2D} \sqrt{N/a_\perp^2}$, where $a_\perp =
\sqrt{\hbar/m\omega_\perp}$, is the radial magnetic length.  Energies are
given in units of $\hbar\omega_\perp$.  We show in Appendix~\ref{app:2d}
that the transversal wave function $\psi_\text{2D}$ fulfills the
following equation
\begin{widetext}
\begin{subequations}\label{eq:psi_2d}
  \begin{align}
    \im \partial_t \psi_\text{2D}(x, y, t) &= \biggl\{ -\frac{1}{2}
    \nabla^2 + V_\text{2D}(x, y) + \frac{\beta_\text{2D}}{\sqrt{2\pi\gamma}}
    \left[1 - \epsilon_{dd} \left(1 - 3 n_z^2 \right) \right]
    |\psi_\text{2D}(x, y, t)|^2 +
    \Phi_\text{2D}  \biggr\} \psi_\text{2D}(x, y, t), \label{eq:psi_2d_a}\\
    \Phi_\text{2D} &= - \frac{3\beta_\text{2D}\epsilon_{dd}}{2}
    \left[\partial_{\bvec n_\perp \bvec n_\perp} - n_z^2 \nabla^2
    \right] \int \di x' \di y' U_\text{2D}(x-x', y-y')
    |\psi_\text{2D}(x', y', t)|^2.\label{eq:phi_2d}
  \end{align}
\end{subequations}
\end{widetext}
Here, $V_\text{2D}(x, y) = (x^2 + y^2)/2$ and $\beta_\text{2D} = 4\pi
N a_s/a_\perp$.  We denote with $\partial_{\bvec n_\perp} = n_x \partial_x
+ n_y \partial_y$ and $\partial_{\bvec n_\perp \bvec n_\perp}
= \partial_{\bvec n_\perp} (\partial_{\bvec n_\perp})$.  The kernel
$U_\text{2D}$ is radially symmetric and is given by
\begin{equation}\label{eq:U_2D}
  U_\text{2D}(r) = \frac{\eu^{r^2/4\gamma}}{(2\pi)^{3/2} \sqrt\gamma}
  K_0(r^2/4\gamma),
\end{equation}
where $K_\nu$ ($\nu$ real) denotes a modified Bessel function of the
second kind and $r^2 = (x-x')^2 + (y-y')^2$.  In
Appendix~\ref{app:potentials} we show that the nonlocal potential,
Eq.~\eqref{eq:phi_2d}, can be written as a simple convolution
$\Phi_\text{2D} = -\tfrac{3\beta_\text{2D}\epsilon_{dd}}{2} \int dx'
dy' U_\text{2D}^{(\bvec n)}(x-x', y-y') |\psi_\text{2D}(x', y',
t)|^2$.  There we also derive a closed form for $U_\text{2D}^{(\bvec
  n)}$, which explicitly depends on the polarization axis.  Assuming
validity of the \GPE, the only approximation in the derivation of
Eq.~\eqref{eq:psi_2d} is the factorization
Eq.~\eqref{eq:factorization_2d}.

In Fig.~\ref{fig:U}(b) we plot the kernel $U_\text{2D}$,
Eq.~\eqref{eq:U_2D}, for different trap anisotropies $\gamma$.  In
contrast to the equivalent 1D kernel $U_\text{1D}$ in
Fig.~\ref{fig:U}(a), the long range behavior does not depend on
$\gamma$.  In fact, we can show that $U_\text{2D}(r) \sim 1/2\pi r$
for $r \rightarrow \infty$.  This is equivalent to the long-range
behavior of the 3D kernel $U_\text{3D}$.  However, in the opposite
limit, $r \rightarrow 0$, we find that the divergence of the kernel is
only logarithmic, $U_\text{2D}(r) \simeq \tfrac{1}{\sqrt{2\pi^3\gamma}}
[-\ln(r) + \ln(2\sqrt\gamma) + \text{const}]$.

For numerical computations in Fourier space, the expression
Eq.~\eqref{eq:phi_2d} for the nonlocal potential is often more useful
than the closed form derived in Appendix~\ref{app:potentials}.
Moreover, in the limit of large trap anisotropy, $\gamma \ll 1$ we
have shown that the potential in Eq.~\eqref{eq:phi_2d} is equivalent
to a Poisson-type equation.  To this end we introduce the fictitious
potential $\phi_\text{2D}$ defined by $\Phi_\text{2D} =
-\frac{3\beta_\text{2D}\epsilon_{dd}}{2}[\partial_{\bvec n_\perp \bvec
  n_\perp} - n_z^2\nabla^2] \phi_\text{2D}$.  For $\gamma \ll 1$ we
may then replace Eq.~\eqref{eq:phi_2d} by
\begin{equation}
  (-\nabla^2)^{1/2} \phi_\text{2D}(x, y, t) =
  |\psi_\text{2D}(x, y, t)|^2.
\end{equation}
Hence, the computation of the nonlocal potential $\Phi_\text{2D}$ in
Fourier space involves only multiplications of the density
$|\psi_\text{2D}|^2$ with the momentum.

In contrast to the 1D mean-field equation, Eq.~\eqref{eq:psi_1d}, in
2D the dipolar interaction increases the contact interaction strength
for dipoles aligned along the $z$ axis (and positive $\epsilon_{dd}$).
This is a manifestation of the fact that magnets aligned in parallel
repel each other.  The modification of the contact interaction term by
a factor of $1/\sqrt{2\pi\gamma}$ is due to the compression along the
$z$ axis~\cite{PetShlWal00,BaoJakMar03}.  Furthermore, unlike in 1D
the effect of dipolar interactions does not vanish at the magic angle
$\tilde\theta_m$: while the dipolar contact interaction term vanishes,
the nonlocal term [last term in Eq.~\eqref{eq:psi_2d_a}] does not.

Analogous to the quasi-1D case, we now derive an approximate
analytical expression for the density.  To this end we assume that a
repulsive contact interaction term (third term) in
Eq.~\eqref{eq:psi_2d_a} dominates the ground state solution.  Hence,
we neglect the kinetic and nonlocal terms in Eq.~\eqref{eq:psi_2d_a}.
With the stationary ansatz $\psi_\text{2D}(x, y, t) = \eu^{-\im\mu_r
  t} \sqrt{n_\text{2D}(r)}$ we find the density profile
$n_\text{2D}(r) = [\mu_r - (r^2/2)] \{\beta_\text{2D}[1 -
\tfrac{\epsilon_{dd}}{\sqrt{2\pi\gamma}} (1 - 3n_z^2)]\}^{-1}$, where
$r^2 = x^2 + y^2$ and $\mu_r$ is the radial part of the chemical
potential.  The density vanishes for $r \geq R = \sqrt{2\mu_r}$.  By
evaluating the normalization of the density, $2\pi \int \di r r
n_\text{2D}(r) = 1$, we find the mean-field radius
\begin{equation}
  R = \left( \frac{4\beta_\text{2D}}{\pi\sqrt{2\pi\gamma}} [1 -
    \epsilon_{dd} (1 - 3n_z^2)] \right)^{1/4}.
\end{equation}
Inserting this radius into the analytical density profile yields
\begin{equation}\label{eq:n_2d}
  n_\text{2D}(r) = \frac{2}{\pi R^2} \left( 1 - \frac{r^2}{R^2}
  \right).
\end{equation}
For the special case $n_z = 1$ our expression for the density
$n_\text{2D}(r)$ coincides with the expression for the ``2D mean-field
regime'' given in Ref.~\cite{ParODe08}.  We can formally generalize
the condition for the validity of Eq.~\eqref{eq:n_2d} given in
Refs.~\cite{ParODe08,MunDel08} to
\begin{equation}\label{eq:condition_2d}
  \frac{\beta_\text{2D}\sqrt{\gamma^3}}{4\pi} [1
  - \epsilon_{dd} (1 - 3n_z^2)] \ll 1.
\end{equation}
While this condition may suggest that $n_\text{2D}(r)$ is a good
approximation for large dipole moment and small axial polarization
$n_z$, we note that in the regime $n_z \alt 1/\sqrt{3}$ the anisotropy
and magnitude of the potential $\Phi_\text{2D}$ increases appreciable.
This may be seen by evaluating the kernel $U_\text{2D}^{\bvec n}$
given in Appendix~\ref{app:potentials}.  In other words,
$n_\text{2D}(r)$ is a good approximation for the 2D density profile if
condition Eq.~\eqref{eq:condition_2d} is fulfilled and the dipoles are
polarized \emph{predominantly} in the axial direction.

By numerically solving Eq.~\eqref{eq:psi_2d} we obtain radial density
profiles of a quasi-2D dipolar \BEC for various trap anisotropies and
polarizations.  For axial polarization ($n_z=1$) we find a radially
symmetric density.  For nonaxial polarization ($n_z<1$) we find that
the quasi-2D \BEC is elongated along the polarization axis projected
onto the $x$-$y$ plane and compressed orthogonal to the polarization
axis.  This is in contrast to the quasi-1D case where the attraction
between aligned dipoles compresses the \BEC along the polarization
axis.  This is a result of the saddle shape of the potential
$U_\text{2D}^{\bvec n}$ with minima along the projection of the dipole
axis.  The experiments in Ref.~\cite{StuGriKoc05,LahKocFro07} show
such an elongation of a dipolar \BEC in the Thomas-Fermi regime.

\begin{figure}
  \centering
  \includegraphics[width=\linewidth]{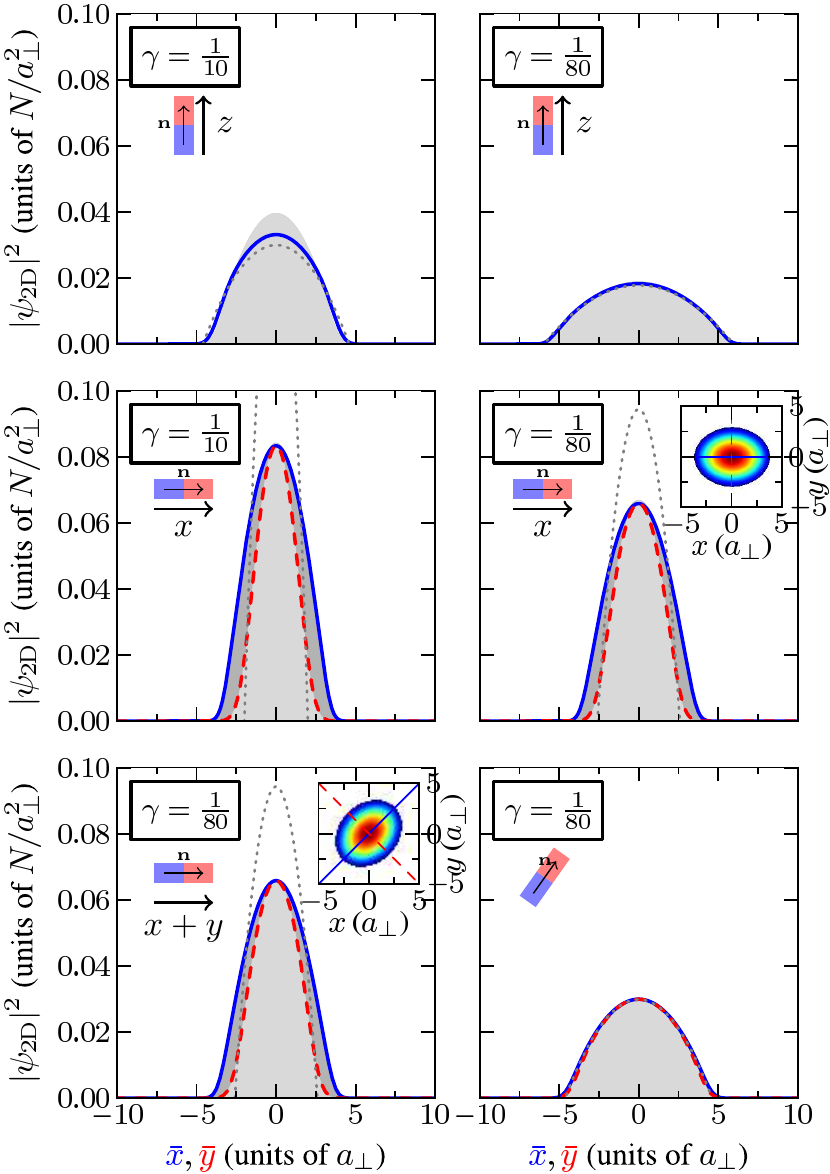}
  \caption{(Color online) Cuts through the radial density profiles of
    the quasi-2D dipolar \BEC given by Eq.~\eqref{eq:psi_2d} for
    various polarizations and trap anisotropies.  The cuts are taken
    along the axes with largest ($\bar x$ axis, solid blue lines) and
    smallest extend of the \BEC ($\bar y$ axis, dashed red).  The
    insets show density plots of the quasi-2D \BEC and the lines
    indicate the position of the cuts ($\bar x$ and $\bar y$ axes,
    respectively). The gray dotted lines are the analytical profiles
    $n_\text{2D}(r)$ and the shaded areas are the profiles obtained
    from the 3D \GPE, Eq.~\eqref{eq:GPE}.  For sufficiently large
    confinement the 3D \GPE profiles are not distinguishable from our
    2D solution.  We choose $\beta_\text{2D} = 100$, $\epsilon_{dd} =
    0.9$ and the dipole axis $\bvec n = (0, 0, 1)$ (top panel), $\bvec
    n = (1, 0, 0)$ (middle panel), $\bvec n = \tfrac{1}{\sqrt
      2}(1,1,0)$ (bottom left panel) and $\bvec n = \tfrac{1}{\sqrt
      3}(1,1,1)$ (bottom right panel).}\label{fig:density_2d}
\end{figure}

In Fig.~\ref{fig:density_2d} we show density profiles of the quasi-2D
\BEC along the elongated (solid blue lines) and compressed axes
(dashed red) in the $x$-$y$ plane.  If the dipoles are aligned
parallel to the symmetry axis of the quasi-2D \BEC ($n_z=1$), the
overall dipolar interaction is repulsive.  This is manifest in
Eq.~\eqref{eq:psi_2d}, where the contact interaction term becomes
larger for positive dipole strength $\epsilon_{dd}$ and the nonlocal
potential is positive.  Moreover, the \BEC remains radially symmetric
as a result of the vanishing radially asymmetric derivative
$\partial_{\bvec n_\perp \bvec n_\perp}$.  We plot the radially
symmetric density profile for $n_z=1$ in the top panel of
Fig.~\ref{fig:density_2d}.  The radial profile of the \BEC becomes
increasingly asymmetric as we move the polarization away from the $z$
axis.  This is evident in the different widths of the density profiles
along the two orthogonal axes in the middle and bottom panels of
Fig.~\ref{fig:density_2d}.  The plots with polarizations along the $x$
axis or the diagonal of the $x$-$y$ plane (middle and bottom left
panel of Fig.~\ref{fig:density_2d}) show the largest difference in
width.  The case of equal polarization in all directions (bottom right
panel in Fig.~\ref{fig:density_2d}), $\bvec n = \tfrac{1}{\sqrt
  3}(1,1,1)$, is special because the dipole interaction does not have
a local character.  This is manifest in Eq.~\eqref{eq:psi_2d} where
the contribution of the dipole interaction to the contact interaction
vanishes at the angle $n_z = \cos\tilde\theta \approx 54.7^\circ$.
Consequently, in the bottom right panel of Fig.~\ref{fig:density_2d}
we only observe a very small asymmetry of the radial \BEC density,
which is a purely nonlocal effect caused by the potential
$\Phi_\text{2D}$.

For comparison, in Fig.~\ref{fig:density_2d} we also plot the density
profiles of a dipolar \BEC obtained by numerically solving the 3D
\GPE, Eq.~\eqref{eq:GPE}, and integrating over the $z$ direction.  We
find excellent agreement with the solutions of our proposed 2D
equations \eqref{eq:psi_2d} for sufficiently large trap anisotropies.
In the top left panel of Fig.~\ref{fig:density_2d} we notice a slight
discrepancy to the 2D solution because the trap anisotropy is not
sufficient to suppress the change in the axial density profile caused
by the dipole interaction.  Furthermore, in Fig.~\ref{fig:density_2d}
we plot the approximation $n_\text{2D}(r)$, Eq.~\eqref{eq:n_2d}.  We
observe good agreement within its regime of validity discussed below
Eq.~\eqref{eq:condition_2d}.  The analytical approximation of the
radial profile agrees well with the numerical ground state solution if
the polarization is predominantly perpendicular to the \BEC disc (top
and bottom right panels of Fig.~\ref{fig:density_2d}).  For
polarizations predominantly in the plane of the quasi-2D \BEC our
proposed Eq.~\eqref{eq:psi_2d} remains a good approximation while the
analytical approximation becomes invalid.

\begin{figure}
  \centering
  \includegraphics[width=\linewidth]{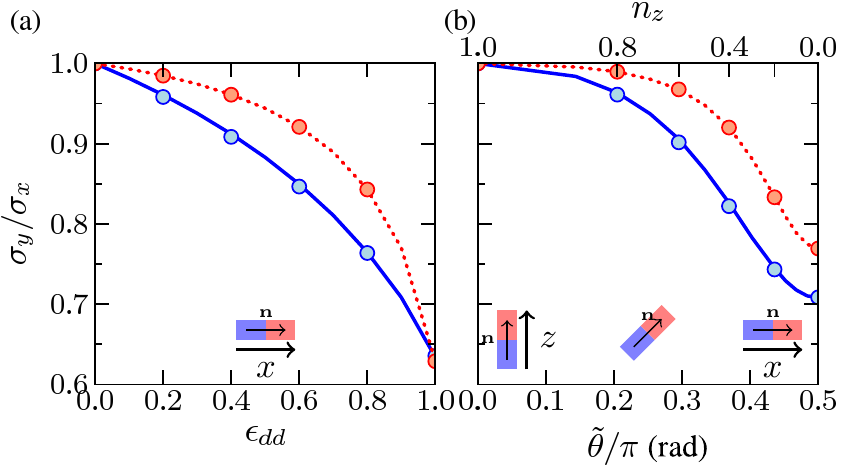}
  \caption{(Color online) Aspect ratio of the quasi-2D \BEC for (a)
    varying dipole strength $\epsilon_{dd}$ with polarization along
    the $x$ axis and (b) varying polarization angle in the $x$-$z$
    plane [$\bvec n = (\sin\tilde\theta, 0, \cos\tilde\theta)$] with
    $\epsilon_{dd} = 0.9$.  We use the trap aspect ratios $\gamma =
    1/10$ (solid lines) and $\gamma = 1/80$ (dotted) with
    $\beta_\text{2D} = 100$.  The circles indicate the corresponding
    condensate aspect ratio according to the numerical solution the 3D
    \GPE, Eq.~\eqref{eq:GPE}.  The upper axis in (b) shows $n_z =
    \cos\tilde\theta$.}\label{fig:aspect}
\end{figure}

We have seen in Fig.~\ref{fig:density_2d} that the quasi-2D \BEC in a
radially symmetric trap loses its radial symmetry if the dipole axis
does not point along its symmetry axis.  This is a consequence of the
anisotropic nature of the dipolar interaction.  It is possible to
observe this effect experimentally by measuring the aspect ratio
\begin{equation}
  \frac{\sigma_y}{\sigma_x} = \frac{\sqrt{\langle y^2
      \rangle}}{\sqrt{\langle x^2\rangle}}
\end{equation}
of the \BEC.  The aspect ratio is particularly suited for measurements
because it is not sensitive on the exact number of
particles~\cite{GioGoePfa03}.

We have computed the aspect ratio of a dipolar \BEC by numerically
solving our 2D equations \eqref{eq:psi_2d} for different values of the
dipole moment and polarization angle.  Fig.~\ref{fig:aspect}
summarizes these results.  In Fig.~\ref{fig:aspect}(a) we plot the
aspect ratio for polarization along the $x$ axis and varying dipole
interaction strength $\epsilon_{dd}$.  We observe that the \BEC
becomes increasingly elongated as we increase the dipolar interaction
strength.  In Fig.~\ref{fig:aspect}(b) we fix the dipole strength but
let the dipolar axis rotate in the $x$-$z$ plane.  This shows how the
radial BEC density profile changes from a symmetric disc to an oval
shape as we rotate the polarization away from the $z$ axis.  We note
that large changes in the aspect ratio occur in a region around the
magic angle $\tilde\theta \simeq \tilde\theta_m$, where the effective
contact interaction is nearly independent of the dipole interaction
strength $\epsilon_{dd}$.  Large trap anisotropies [dashed line in
Fig.~\ref{fig:aspect}(b)] suppress the onset of significant \BEC
asymmetry because the contact interaction dominates the ground state.
We have also obtained the aspect ratios by numerically solving the 3D
\GPE, Eq.~\eqref{eq:GPE} and integrating over the $z$ axis (circles in
Fig.~\ref{fig:aspect}).  We find excellent agreement with the aspect
ratios obtained from our 2D equations.  Since a rotation of the
polarization in the $x$-$y$ plane only corresponds to a rotation of
the elongated axis of the ground state density (see
Fig.~\ref{fig:density_2d}), we obtain similar results to
Fig.~\ref{fig:aspect}(b) for arbitrary polarization.  This shows that
the reduced Eqs.~\eqref{eq:psi_2d} are indeed a good approximation for
describing quasi-2D dipolar \BEC{}s at arbitrary polarization with
sufficiently strong axial trapping.

\section{Conclusion}\label{sec:conclusion}
We have presented Gross-Pitaevskii-type mean-field equations for
trapped quasi-1D, Eq.~\eqref{eq:psi_1d}, and quasi-2D,
Eq.~\eqref{eq:psi_2d}, dipolar \BEC{}s polarized along an arbitrary
axis.  These equations are based on a rigorous dimension reduction of
the full 3D \GPE.  In contrast to previous works, they are valid for
arbitrary dipole alignment in the mean-field regime if the \BEC is in
the ground state of the radial or axial harmonic trap, respectively.
Our result shows that quasi-1D and quasi-2D dipolar \BEC{}s are
governed by a modified contact interaction term and an additional
nonlocal potential.  We have given explicit expressions for the
nonlocal potential for arbitrary polarization (also see
Appendix~\ref{app:potentials}).

One of the main advantages of the proposed mean-field equations is
that they are well suited for numerical computations in strongly
confined \BEC{}s.  Our numerical implementations of the ground state
computation in 1D and 2D perform much faster than our equivalent 3D
computations.  This is because we only need to integrate over the
reduced dimensions, which vary over similar time scales, whereas the
excluded dimensions vary on a much faster time scale for strong trap
anisotropies.  Moreover, the kernel of the convolution in the nonlocal
potential is bounded in 1D and diverges only logarithmically in 2D.
In contrast, in 3D the corresponding kernel diverges inverse linearly.
Our formulation of the nonlocal potential in terms of partial
derivatives allows for efficient numerical methods based on the
Fourier transformation.

We have computed the ground states of our 1D and 2D equations
numerically and compared them with the ground states of the 3D \GPE.
We find excellent agreement but notice small discrepancies in 1D for
the case when the \BEC is polarized perpendicular to the elongated
direction.  By neglecting the kinetic energy term and assuming a
vanishing nonlocal potential we have derived analytical expressions
for the density profiles of quasi-1D and quasi-2D dipolar \BEC{}s when
the dipoles are aligned predominantly along the $z$ axis.  The ground
state of the quasi-2D dipolar \BEC becomes anisotropic if the
polarization is not parallel to the symmetry axis of the
pancake-shaped \BEC.  This results in a varying aspect ratio of the
\BEC pancake, which can be measured with time-of-flight imaging.  We
have computed this aspect ratio for varying dipole interaction
strength, trap anisotropy and polarization axis.

As an outlook, we point out that it is straightforward to use our
equations for studying the dynamics of arbitrarily polarized dipolar
\BEC{}s in lower dimensions.  They also allow for a stability analysis
of such lower dimensional dipolar \BEC{}s.  In a future work we will
investigate the influence of a changing dipole axis on the formation
of vortices in 2D dipolar \BEC{}s in a rotating frame.

\begin{acknowledgments}
  We would like to thank Hanquan Wang for his help with some of the
  numerical computation.  This work was supported by the Academic
  Research Fund of Ministry of Education of Singapore grant
  R-146-000-120-112.  ZL acknowledges support in part by National
  Science Foundation of China (NSFC) grants 10801029 and 10911120384,
  FANEDD, Shanghai Rising Star Program (10QA1400300), SGST
  09DZ2272900, and SRF for ROCS-SEM.
\end{acknowledgments}

\appendix

\section{Derivation of the 1D mean-field equation}\label{app:1d}
Under the assumption in Sec.~\ref{sec:1d}, plugging $\psi(\bvec r, t)
= \eu^{-\im\omega_\perp t} w_\text{2D}(x, y) \psi_\text{1D}(z, t)$ into the \GPE,
Eq.~\eqref{eq:psi_3d_a}, after rescaling, we have
\begin{equation}
  \im\partial_t \psi_\text{1D}w_\text{2D}= \left(
    \frac{ -\partial_{zz} + z^2}{2} +
    \beta_1|\psi_\text{1D}w_\text{2D}|^2 +
    \partial_{\bvec n \bvec n}\Phi
  \right) \psi_\text{1D}w_\text{2D},\label{Append:psi_1d}
\end{equation}
where
\begin{equation}
  \Phi = -3\beta_\text{1D}\epsilon_{dd}  \int d^3 \bvec r'
  U_\text{3D}(\bvec r - \bvec r')
  |(\psi_\text{1D}w_\text{2D})(\bvec r', t)|^2\label{Append:Phi_1d}
\end{equation}
with the rescaled quantities $w_\text{2D}(x,y) =
\sqrt{\frac{\gamma}{\pi}} e^{-\frac{\gamma}{2}(x^2 + y^2)}$, $\beta_1
= \beta_\text{1D} (1 - \epsilon_{dd})$.  By multiplying both sides of
Eq.~\eqref{Append:psi_1d} by $w_\text{2D}(x,y)$ and integrating over
the $x$-$y$ plane we obtain an equation in $z$ only.  Since
$w_\text{2D}(x,y)$ is normalized according to $\int dxdy
w_\text{2D}^2(x,y) = 1$ and $\int dxdy w_\text{2D}^4(x,y) =
\gamma/2\pi$, this integration is straightforward for all but the last
term.  In the following we outline the calculation of this last term.

In order to obtain Eq.~\eqref{eq:psi_1d}, we need to compute
\begin{equation}\label{eq:app:dnn-Phi_1d}
  \iint_{-\infty}^\infty  \di x\di y \partial_{\bvec n\bvec
    n} \Phi(x,y,z,t) w_\text{2D}^2(x,y).
\end{equation}
Noticing the symmetry of $U_\text{3D}$ and $w_\text{2D}$ in $x$ and $y$ we see
that
\begin{equation}\label{eq:app:U-symm}
  \partial_{xx} \left(U_\text{3D}*
    |\psi_\text{1D}w_\text{2D}|^2\right)=\partial_{yy}\left(U_\text{3D}*
    |\psi_\text{1D}w_\text{2D}|^2\right),
\end{equation}
where we denote with $a*b$ the convolution of $a$ and $b$.  Since
$U_\text{3D}$ is the Green's function of the Poisson
equation~\cite{ODeGioEbe04}, we have $-\nabla^2 U_\text{3D}(\bvec r) =
\delta(\bvec r)$.  Inserting Eq.~\eqref{eq:app:U-symm} into this
expression yields
\begin{equation}\label{eq:app:dxx-U_3d}
  \begin{split}
    \partial_{xx} \left(U_\text{3D}*
      |\psi_\text{1D}w_\text{2D}|^2\right) &= -\frac{|\psi_\text{1D}
      w_\text{2D}|^2}{2}\\
    &\quad - \frac{\partial_{zz} \left( U_\text{3D} * |\psi_\text{1D}
        w_\text{2D}|^2 \right)}{2}.
  \end{split}
\end{equation}
Moreover, $|\psi_\text{1D}w_\text{2D}|^2$ is an even function with respect to
$x$ and $y$ and so is $U_\text{3D}* |\psi_\text{1D}w_\text{2D}|^2$, which
implies that the partial derivatives $\partial_{xy}$, $\partial_{yz}$,
$\partial_{xz}$ of $U_\text{3D}* |\psi_\text{1D}w_\text{2D}|^2$ are odd
functions in $x$ and $y$.  Recalling $\partial_{\bvec n\bvec n} =
n_x^2\partial_{xx} + n_y^2\partial_{yy} + n_z^2\partial_{zz} +
2n_xn_y\partial_{xy} + 2n_yn_z\partial_{yz} + 2n_xn_z\partial_{xz}$ we
plug Eq.~\eqref{eq:app:dxx-U_3d} into Eq.~\eqref{eq:app:dnn-Phi_1d},
which yields
\begin{multline}\label{eq:app:dnn-Phi_1d-red}
  3\beta_\text{1D}\epsilon_{dd} \bigg(\frac{1-n_z^2}{2}
  \frac{\gamma}{2\pi} |\psi_\text{1D}|^2\\
  \quad+ \frac{1 - 3n_z^2}{2} \partial_{zz} \iint \di x \di y \left(
    U_\text{3D} * |\psi_\text{1D}w_\text{2D}|^2 \right) w_\text{2D}^2
  \bigg)
\end{multline}
because integrals of odd functions (the mixed partial derivatives)
vanish.  We see that the first term in
Eq.~\eqref{eq:app:dnn-Phi_1d-red} contributes to the modification of
the contact interaction term in Eq.~\eqref{eq:psi_1d_a}.

In order to evaluate the last term in
Eq.~\eqref{eq:app:dnn-Phi_1d-red} we expand the convolution and
compute the resulting integrals
\begin{equation*}
  \int_{\mathbb{R}^4} \di x'\di y'\di x\di y
  \frac{w_\text{2D}^2(x^\prime,y^\prime)
    w_\text{2D}^2(x,y)}{4\pi\sqrt{(x - x^\prime)^2 +
      (y - y^\prime)^2 + (z - z^\prime)^2}}.
\end{equation*}
To this end, we introduce the variables $\tilde{x} = x-x^\prime$,
$\tilde{x}^\prime = x+x^\prime$, $\tilde{y} = y-y^\prime$,
$\tilde{y}^\prime = y+y^\prime$ and integrate over $\tilde x^\prime$
and $\tilde y^\prime$.  The result is
\begin{equation}\label{eq:app:int-R2}
  \iint \di\tilde x\di\tilde y \frac{\gamma e^{-\frac{\gamma}{2}
      (\tilde{x}^2 + \tilde{y}^2)}}{8\pi^2\sqrt{\tilde{x}^2 +
      \tilde{y}^2 + (z-z^\prime)^2}},
\end{equation}
which can be evaluated further in polar coordinates $\tilde x =
r\cos\theta'$, $\tilde y = r\sin\theta'$.
Equation~\eqref{eq:app:int-R2} reduces to
$\tfrac{\sqrt\gamma}{4\sqrt{2\pi}} U_\text{1D}(z-z')$ with the
substitution $\zeta = \sqrt{r^2 + (z-z^\prime)^2}$ and
\begin{equation}\label{eq:app:U_1d}
  U_\text{1D}(z) = \sqrt\frac{2\gamma}{\pi} \eu^{\gamma z^2/2}
  \int_{|z|}^\infty \di \zeta \eu^{-\gamma \zeta^2/2}.
\end{equation}
Equation~\eqref{eq:app:U_1d} is an integral representation of
$U_\text{1D}$ given in Eq.~\eqref{eq:U_1d}.  Inserting
Eq.~\eqref{eq:app:U_1d} back into Eq.~\eqref{eq:app:dnn-Phi_1d-red} we
obtain the potential $\Phi_\text{1D}$, Eq.~\eqref{eq:phi_1d}.
Plugging Eq.~\eqref{eq:app:dnn-Phi_1d-red} into the integrated
Eq.~\eqref{Append:psi_1d} results in the mean-field equation for a
quasi-1D dipolar \BEC, Eq.~\eqref{eq:psi_1d}.

The Fourier transform of $U_\text{1D}(z)$ is given by
\begin{equation}
  \widehat{U_\text{1D}}(k_z) = \frac{1}{\sqrt{\gamma}\pi} \int_0^\infty
  \di \kappa \frac{e^{-\frac{\kappa}{2\gamma}}}{k_z^2 + \kappa}.
\end{equation}
The asymptotic behavior of this Fourier transform is
$\widehat{U_\text{1D}}(k_z) = \tfrac{1}{\sqrt\gamma\pi} [\ln(2\gamma)
- \gamma_e - 2\ln|k_z|] + \mathcal{O}(k_z)$ for $|k_z|\rightarrow 0$
and $\widehat{U_\text{1D}}(k_z) \sim \tfrac{2\sqrt\gamma}{\pi}
\tfrac{1}{|k_z|^2}$ for $ |k_z|\rightarrow \infty$, where $\gamma_e$
is the Euler-Mascheroni constant.

\section{Derivation of the 2D mean-field equation}\label{app:2d}
Under the assumptions in Sec.~\ref{sec:2d}, plugging $\psi(\bvec r, t)
= \eu^{-\im\omega_z t/2} w_\text{1D}(z) \psi_\text{2D}(x,y, t)$ into the \GPE,
Eq.~\eqref{eq:psi_3d_a}, after rescaling, we have
\begin{equation}
  \im\partial_t \psi_\text{2D}w_\text{1D} = \left\{
    \frac{ -\nabla_\perp^2 + V_\text{2D}}{2} +
    \beta_2|\psi_\text{2D}w_\text{1D}|^2 +
    \partial_{\bvec n \bvec n}\Phi
  \right\} \psi_\text{2D}w_\text{1D},\label{Append:psi_2d}
\end{equation}
where now
\begin{equation}
  \Phi =-3\beta_\text{2D}\epsilon_{dd}  \int d^3 \bvec r' U_\text{3D}(\bvec
  r - \bvec r') |(\psi_\text{2D}w_\text{1D})(\bvec r', t)|^2,\label{Append:Phi_2d}
\end{equation}
with rescaled $w_\text{1D}(z)=\left(\frac{1}{\gamma\pi}\right)^{1/4}e^{-\frac{
    z^2}{2\gamma}}$, $\nabla^2_\perp=\partial_{xx}+\partial_{yy}$,
$\beta_2 = \beta_\text{2D}(1-\epsilon_{dd})$.  Multiplying both sides
of Eq.~\eqref{Append:psi_2d} by $w_\text{1D}(z)$ and integrating over the $z$
direction, we can obtain a 2D wave equation for $\psi_\text{2D}$.

As in the preceding appendix, the integration is straightforward for
all but the last term because $\int dz w_\text{1D}^2(z) = 1$ and $\int
dz w_\text{1D}^4(z) = 1/\sqrt{2\pi\gamma}$.  In the following we only
present the integration of the last term of Eq.~\eqref{Append:psi_2d},
\begin{equation}\label{app:d_nnPhi_2D}
  \int_{-\infty}^\infty \di z \partial_{\bvec n\bvec n} \Phi(x,y,z,t)
  w_\text{1D}^2(z).
\end{equation}
Again we use the identity $-\nabla^2 U_\text{3D}(\bvec r) =
\delta(\bvec r)$ to write
\begin{equation*}
\begin{split}
  \partial_{zz} \left(U_\text{3D} *
    |\psi_\text{2D}w_\text{1D}|^2\right) &=
  -|\psi_\text{2D} w_\text{1D}|^2\\
  &\quad - \nabla^2_\perp \left( U_\text{3D} * |\psi_\text{2D}
    w_\text{1D}|^2 \right).
\end{split}
\end{equation*}
We recall that $|\psi_\text{2D}(x,y,t) w_\text{1D}(z)|^2$ is even in $z$ so that
$U^{3D} * |\psi_\text{2D} w_\text{1D}|^2$ also becomes even in $z$ and
$\partial_{\bvec n_\perp z} (U^{3D}*|\psi_\text{2D}w_\text{1D}|^2)$ becomes odd
in $z$.  Plugging $\partial_{\bvec n\bvec n} = \partial_{\bvec
  n_\perp\bvec n_\perp}+n_z^2\partial_{zz}+2n_z\partial_{\bvec n_\perp
  z}$ into Eq.~\eqref{app:d_nnPhi_2D} yields
\begin{multline}\label{eq:app:dnn-Phi_2d-red}
  3\beta_\text{2D}\epsilon_{dd} \bigg(
  \frac{n_z^2}{\sqrt{2\gamma\pi}} |\psi_\text{2D}|^2\\
  \quad - (\partial_{\bvec n_\perp\bvec n_\perp} - n_z^2 \nabla^2_\perp)
  \int_{-\infty}^\infty \di z\left( U_\text{3D} * |\psi_\text{2D}
    w_\text{1D}|^2 \right) w_\text{1D}^2(z) \bigg),
\end{multline}
where the $\partial_{\bvec n_\perp z}$ term disappears because the
integral of odd functions vanishes.  The first term in
Eq.~\eqref{eq:app:dnn-Phi_2d-red} contributes to the modified contact
interaction term in Eq.~\eqref{eq:psi_2d}.

After expanding the convolution in the last term of
Eq.~\eqref{eq:app:dnn-Phi_2d-red}, wee see that we need to compute the
integral
\begin{equation}\label{eq:app:convolution}
  \iint_{-\infty}^\infty \di z' \di z \frac{w_\text{1D}^2(z^\prime)
    w_\text{1D}^2(z)}{4\pi\sqrt{(x-x^\prime)^2 + (y-y^\prime)^2 +
      (z-z^\prime)^2}}.
\end{equation}
By changing the variables according to $\tilde{z}=z-z^\prime$,
$\tilde{z}^\prime=z+z^\prime$, and integrating over $\tilde z'$ we reduce
Eq.~\eqref{eq:app:convolution} to
\begin{equation}\label{eq:int-R}
  \int_{-\infty}^\infty \di\tilde z
  \frac{1}{2 \sqrt\gamma (2\pi)^{3/2}}\frac{e^{-\tilde z^2/2\gamma}}{\sqrt{(x -
      x^\prime)^2 + (y-y^\prime)^2 + \tilde z^2}}.
\end{equation}
By substituting $\rho = \tilde z/\sqrt\gamma$ and introducing polar
coordinates with $r = \sqrt{(x-x')^2 + (y-y')^2}$ we get
\begin{equation}\label{eq:app:U_2d}
  U_\text{2D}(r) = \frac{1}{(2\pi)^{3/2}} \int_{-\infty}^\infty \di \rho
  \frac{\eu^{-\rho^2/2}}{\sqrt{r^2 + \gamma \rho^2}}.
\end{equation}
This integral representation of $U_\text{2D}(r)$ is identical to
Eq.~\eqref{eq:U_2D}~\cite[see Sec. 3.364]{GraRyz00}.  We obtain the
potential $\Phi_\text{2D}$, Eq.~\eqref{eq:phi_2d}, from the last term
in Eq.~\eqref{eq:app:dnn-Phi_2d-red} after inserting
Eq.~\eqref{eq:app:U_2d}.  Finally, plugging
Eq.~\eqref{eq:app:dnn-Phi_2d-red} into the integrated
Eq.~\eqref{Append:psi_2d} results in the mean-field equation for the
quasi-2D dipolar BEC, Eq.~\eqref{eq:psi_2d}.

The Fourier transform $\widehat{U_\text{2D}}(k_x, k_y)$ of the
radially symmetric kernel $U_\text{2D}$ is also radially symmetric and
given by
\begin{equation}
  \widehat{U_\text{2D}}(|k_r|) = \frac{1}{2\pi^2} \int_{-\infty}^\infty
  \di \kappa \frac{e^{-\frac{\gamma \kappa^2}{2}}}{|k_r|^2 + \kappa^2}
\end{equation}
with $|k_r| = \sqrt{k_x^2 + k_y^2}$.  The asymptotic behavior of this
Fourier transform is $\widehat{U_\text{2D}}(|k_r|) \sim
\frac{1}{2\pi |k_r|}$ for $|k_r|\rightarrow 0$ and
$\widehat{U_\text{2D}}(|k_r|) \sim \frac{1}{\sqrt{2\pi^3\gamma}}
\frac{1}{|k_r|^2}$ for $|k_r|\rightarrow\infty$.

\section{Closed forms of the 1D and 2D nonlocal
  potentials}\label{app:potentials}
In this appendix, we derive closed forms for the nonlocal potentials
$\Phi_\text{1D}$, Eq.~\eqref{eq:phi_1d}, and $\Phi_\text{2D}$,
Eq.~\eqref{eq:phi_2d}.  Owing to the properties of the convolution in
$\Phi_\text{1D}$ we may write Eq.~\eqref{eq:phi_1d} as
\begin{equation*}
  \Phi_\text{1D} =
  \frac{3\beta_\text{1D}\epsilon_{dd}\sqrt\gamma}{8\sqrt{2\pi}} (1 -
  3n_z^2) (\partial_{zz} U_\text{1D}) * |\psi_\text{1D}|^2,
\end{equation*}
\ie, the derivatives only affect the kernel $U_\text{1D}$.
Straightforward calculation of the second derivative of
Eq.~\eqref{eq:U_1d} leads to
\begin{equation*}
  \tilde U_\text{1D}(z) = \partial_{zz} U_\text{1D}(z) = \gamma
  U_\text{1D}(z) (1 + \gamma z^2) - \sqrt{\frac{2\gamma^3}{\pi}}
  |z|.
\end{equation*}
The resulting nonlocal potential $\Phi_\text{1D}$ is of the same form
as the one given in Ref.~\cite{SinSan07} for the special case of axial
polarization.

In an analogous fashion, $\Phi_\text{2D}$ in Eq.~\eqref{eq:phi_2d} is
given by
\begin{equation}\label{eq:Phi_2D_2}
  \Phi_\text{2D} = -\frac{3\beta_\text{2D}\epsilon_{dd}}{2} \left[
    \left( \partial_{\bvec n_\perp \bvec n_\perp} - n_z^2\nabla^2
    \right) U_\text{2D} \right] * |\psi_\text{2D}|^2.
\end{equation}
By using Eq.~\eqref{eq:U_2D} and the properties of the derivatives of
Bessel functions we find for the second derivatives of $U_\text{2D}$
\begin{align*}
  \partial_{xx}^2 U_\text{2D} &=
  \frac{e^{r^2/4\gamma}}{2(2\pi\gamma)^{\tfrac{3}{2}}} \left[\biggl( 1 +
    \frac{x^2}{\gamma} \right) K_0 - \left(1 + \frac{x^2}{\gamma}
    - \frac{2x^2}{r^2} \right) K_1 \biggr],\\
  \partial_{yy}^2 U_\text{2D} &=
  \frac{e^{r^2/4\gamma}}{2(2\pi\gamma)^{\tfrac{3}{2}}}
  \biggl[\left( 1 + \frac{y^2}{\gamma} \right) K_0 - \left(1 +
    \frac{y^2}{\gamma} - \frac{2y^2}{r^2} \right) K_1 \biggr],\\
  \partial_{xy}^2 U_\text{2D} &= \partial_{yx}^2 U_\text{2D} =
  \frac{e^{r^2/4\gamma}}{2(2\pi\gamma)^{\tfrac{3}{2}}} \biggl[
  \frac{xy}{\gamma} K_0 - \left( \frac{xy}{\gamma} -
    \frac{2xy}{r^2} \right) K_1 \biggr].
\end{align*}
Here and in the following we suppress the argument $r^2/4\gamma$ of
the Bessel functions with $r^2 = x^2 + y^2$.  Plugging these
identities into Eq.~\eqref{eq:Phi_2D_2} we find for the nonlocal
kernel $U_\text{2D}^{(\bvec n)}(x, y) = (\partial_{\bvec n_\perp \bvec
  n_\perp} - n_z^2\nabla^2) U_\text{2D}(r)$
\begin{widetext}
  \begin{equation}\label{eq:U_2D_tilde}
    U_\text{2D}^{(\bvec n)}(x, y) =
    \frac{e^{r^2/4\gamma}}{2(2\pi\gamma)^{3/2}}
    \biggl[ \left( 1 -
      3n_z^2 + \frac{(n_x x + n_y y)^2 - n_z^2 r^2}{\gamma}
    \right) K_0
    - \left( 1 - n_z^2 + \frac{(n_x x + n_y y)^2 [1 -
        2\gamma/r^2] - n_z^2 r^2}{\gamma} \right) K_1 \biggr]
  \end{equation}
\end{widetext}
and the nonlocal potential $\Phi_\text{2D} =
-\tfrac{3\beta_\text{2D}\epsilon_{dd}}{2} U_\text{2D}^{(\bvec n)} *
|\psi_\text{2D}|^2$.  The kernel $U_\text{2D}^{(\bvec n)}(x, y)$
changes from a symmetric peak for $n_z=1$ to a saddle shape with
minima along the projection of the dipole axis onto the $x$-$y$ plane
for $n_z < 1$.  We note that $\Phi_\text{2D}$ reduces to the potential
for axial polarization ($n_z=1$) derived, \eg, in
Ref.~\cite{KomCoo07}.  However, Eq.~\eqref{eq:U_2D_tilde} is valid for
arbitrary polarization.

\bibliography{dipolar}

\end{document}